# Colloidal electro-phoresis in the presence of symmetric and asymmetric electro-osmotic flow

Denis Botin*, Jennifer Wenzl[#], Ran Niu[∞], Thomas Palberg

Institute of Physics, Johannes Gutenberg University, D-55099 Mainz, Germany

*corresponding author: dbotin@uni-mainz.de

[#]present address: J. Wenzl, Dept. of Physics, Univ. Koblenz-Landau, D-56016 Koblenz, Germany

[∞]present address: R. Niu, Dept. of Physics, Cornell University, Ithaca, New York, 14853

**Abstract.**

We characterize the electro-phoretic motion of charged sphere suspensions in the presence of substantial electro-osmotic flow using a recently introduced small angle super-heterodyne dynamic light scattering instrument (ISASH-LDV). Operation in integral mode gives access to the particle velocity distribution over the complete cell cross-section. Obtained Doppler spectra are evaluated for electro-phoretic mobility, wall electro-osmotic mobility and particle diffusion coefficient. Simultaneous measurements of differing electro-osmotic mobilities leading to asymmetric solvent flow are demonstrated in a custom made electro-kinetic cell fitting standard microscopy slides as exchangeable sidewalls. Scope and range of our approach are discussed demonstrating the possibility of an internal calibration standard and using the simultaneously measured electro-kinetic mobilities in the interpretation of microfluidic pumping experiment involving an inhomogeneous electric field and a complex solvent flow pattern.

# Introduction

Electro-kinetic properties of particles and surfaces are of key importance in many fields ranging from wet processing of materials over microfluidics to bio-systems [1]. The present paper focuses on the simultaneous measurement of two different electro-kinetic quantities: particle electro-phoretic velocities and substrate electro-osmotic velocities. As first observed by v. Smoluchowski,[2], both rely on the same effects. A homogeneous electric field applied to a colloidal suspension of charged particles induces a slip-motion between their charged surfaces and the adjacent solvent. For a stationary solvent, the resulting particle velocity $v_p$ equals the electro-phoretic slip velocity $v_{ep} = E\,\mu_{ep}$ which depends on field strength E and particle mobility $\mu$. Likewise, the field induces a slip motion between a charged container wall and the adjacent solvent. With the cell kept fixed, the lab frame electro-osmotic velocity is $u_{eo} = E\,\mu_{eo}$ with the electro-osmotic mobility $\mu_{eo}$. In closed containers, the incompressibility of the solvent results in substantial backflows, i.e. the electro-osmotic counter-pressure provokes a solvent flow, $u_s(x,y)$, of Poiseuille type. Clearly, then the particle velocity becomes a function of location, even in a homogeneous electric field: $v_P(x,y) = v_{ep} + u_s(x,y)$. Depending on the electro-kinetic mobility of interest, different strategies have been proposed to isolate the respective quantity.

The traditional standard technique of micro-electro-phoresis measures the complete flow profile in a somewhat tedious microscopy experiment to determine the particle velocity at the so-called stationary level, where the net solvent-flow theoretically vanishes [3]. Also in Laser Doppler Velocimetry (LDV) [4] one typically focuses on the stationary level [3, 5, 6]. Hence electro-osmosis id typically regarded as a nuisance in electro-phoretic experiments. To minimize electro-osmosis, wall coatings e.g. by Bovine Serum Albumin (BSA) have been suggested, but in practice this often generates the problem of particle adsorption at the cell walls and an uncontrolled alteration and distortion of the solvent flow profile [3]. Moreover, also small deviations in the measuring position can give significant errors in the measured mobility, especially if the cell wall is highly charged. Application of sinusoidal or rectangular fast field switch hampers the (full) development of the electro-osmotic solvent flow and yields central plug-flow. Then, analytical modelling allows estimates of $v_{ep}$ from $v_p$ and a known $\mu_{eo}$ [7, 8]. Also alternative cell designs have been studied, e.g. Tiselius-type cells with large bypass [9] as well as cells with small-gap electrodes freely suspended far off any walls [10, 11, 12].

The standard method to measure the electro-osmotic mobility is the streaming potential [1, 3] but also microscopic determinations of suspension flow-profiles in standard or custom-made micro-electro-phoresis cells have been performed [13, 14, 15]. We here adapt the latter approach but combine a suitable electro-kinetic cell with a recently introduced versatile version of LDV [16].

This is further motivated by the fact that in many technical applications and in particular contemporary micro-fluidic experiments both mobilities are needed simultaneously to interpret the experimental findings. A typical examples are micro-fluidic applications like electrophoretic

[17], diffusio-phoretic [18] and osmotic trapping [19] or single particle electro-kinetic experiments of sedimented and optically trapped particles [20, 21, 22, 23 ]. Here often inhomogeneous electric fields are employed which in addition may vary in time. In phoretic micro-swimming and electro-osmotic pumping [24, 25, 26] typical experiments involve locally generated diffusio-electric fields. [27, 28]. Also here we are relying on well known values for the electro-kinetic mobilities of substrates and transported particles. The electro-osmotic mobilities of several high purity materials are well documented in literature. However, for many standard materials like glass, quartz, PMMA or PDMS results show a large spread of values depending on cleaning and conditioning conditions. Moreover, in microfluidics and other applications, coated cell walls are frequently used, which introduces additional dependence of $\mu_{eo}$ on the preparation protocol followed. In this situation, it would be highly desirable to have a fast and reliable characterization method available, which in addition would also yield the particle mobility under exactly the same conditions.

Our integral small angle super-heterodyne dynamic light scattering instrument (ISASH-LDV) was designed to determine the electro-phoretic mobilities of charged colloidal spheres in aqueous suspension in the presence of substantial electro-osmosis. Like Phase Analysis Light Scattering (PALS [12, 29]), the ISASH-LDV instrument works in a super-heterodyne configuration which allows separating the desired heterodyne spectrum. However, we here use the instrument in *integral* mode, collecting the signal over the complete cross-section of the electro-kinetic cell [30]. Thus the super-heterodyne (shet) spectrum is directly proportional to the of particle velocity distribution and obtained in a single measurement. No tedious point by point measurements are needed to check the location of the stationary plane, analyse the data for the electro-phoretic particle velocity and derive the electro-osmotic solvent velocities at the cell walls. Instead, both the electro-phoretic and electro-osmotic velocities can be measured simultaneously and deviations from theoretical expectations (profile distortion, shear banding, field dependent mobilities, etc.) can be easily checked. Moreover, ISASH-LDV collects light at small angles, facilitating studies of the suspension electro-kinetics irrespective of the sample's structure [31, 32]. Adapting the scattering configuration to super-heterodyning allows convenient discrimination between the shet-signal containing the desired informations and the homodyne signal as well as low frequency noise. Small angle shet-scattering has been successfully applied to study electro-phoretic mobilities for colloidal dispersions at different particle volume fractions, electrolyte concentrations including non-interacting systems as well as colloidal fluids and crystals [33, 34]. Finally, in combination with a suitably restricted detection volume and direction, ISASH-DLS allows application of a facile correction procedure to eliminate multiply scattered light. The instrument is thereby capable to work at transmissions as low as 40% [16].

Here we are interested in experiments involving both type of electro-kinetic mobilities. We employed a custom made electro-kinetic cell with exchangeable sidewalls to realize symmetric and asymmetric electro-osmotic flows. The cell can mount differently coated standard microscopy slides and can be connected to a standard conditioning circuit to adjust the particle number density and the electrolyte concentration in a controlled way. Field strength dependent measurements thus facilitate simultaneous measurements of an electro-phoretic particle mobility and two electro-osmotic mobilities. We further demonstrate that this capability can be used to provide an internal mobility standard. Finally, to provide a worked example for the use of both simultaneously determined mobilites, we turn to an interpretation of an electro-osmotic pumping experiment from micro-fluidics that involves charged tracers and differently conditioned substrates [25]. There we can verify that the triggered solvent flux (as inferred from the tracer velocity) depends linearly on the substrate electro-osmotic mobility.

In what follows, we first shortly recall the ISASH-LDV approach [16] and present the underlying theory. Then we will introduce the studied system and experimental setup, focusing on the newly designed electro-kinetic cell with exchangeable sidewalls. We will also shortly introduce the micro-fluidic pump experiment. In the result section, we first present measurements of electro-phoretic and electro-osmotic mobilities for the case of symmetric solvent flows. We then turn to asymmetric flows and introduce the internal standard for electro-osmosis. We close with the discussion of the range and scope of the here detailed ISASH variant of LDV and exemplify it with the interpretation of the mentioned microfluidic pump experiment.

## Experimental

### Small-angle super-heterodyne Dynamic light scattering (SASH-LDV).

The instrument has recently been described and its performance characterized in detail elsewhere [16]. It employs super-heterodyning [35] to separate the desired super-heterodyne part from homodyne and low frequency noise. In short, light scattered by the particles off the Illuminating beam and the transmitted Reference beam (acting as local oscillator) superimpose at the detector. The mixed-field auto-correlation function, $C_{shet}(q,\tau)$, contains terms stemming from the homodyne scattering (scattered light mixing with scattered light) and from the heterodyne scattering (scattered light mixed with local oscillator light). Applying a frequency shift, $\omega_B$, between Reference and Illuminating beam (super-heterodyning), allows separating these contributions in frequency space *via* a shift of the shet-part by $\omega_B$ [31]. ISASH further features small angle scattering, which allows an efficient correction scheme to isolate the single-scattering signal from undesired multiple-scattering contributions [16]. A sketch of the detection of the setup part is given in top view in Fig 1a.

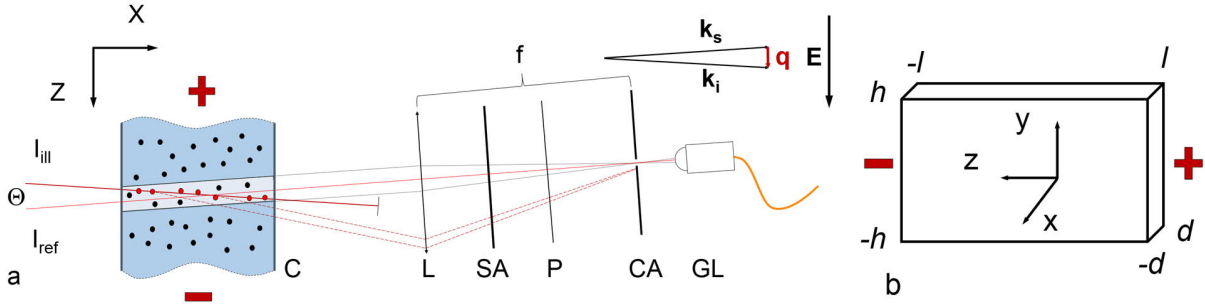

Fig. 1: Sketches of the experimental set-up. (a) Top view of the scattering geometry in the x-z-scattering plane. Illuminating beam ($I_{ill}$, $\omega_{ill} = \omega_0$) and Reference beam ($I_{ref}$, $\omega_{ref} = \omega_0+\omega_{Bragg}$), cross under an angle $\theta$ inside the suspension filled sample cell (C) (refraction at cell walls not drawn here). The Reference beam is collinear with the observation direction. Scattered light is focused by lens (L) at f = 50 mm onto a small circular aperture (CA) and collected by a gradient-index (GRIN) lens (GL) and fed into an optical fiber leading to the detector (not shown). Precise location of CA selects the detected vector of scattered light and restricts it to $\mathbf{k}_f = \mathbf{k}_{ref}$ yielding a scattering vector q parallel to the direction of the applied electric field E (downwards to the cathode). The distance CA-GL defines the diameter of the cylindrical (light grey area inside the cell) adjusted to contain the complete path of the Illuminating beam inside the cell. A horizontal slit aperture (SA) rejects any light travelling outside the x-z plane. A polarizer (P) assures V/V scattering geometry. (b) Cell geometry and coordinates. The cell has a rectangular cross section with height in y-direction 2h, depth in x-direction 2d and length, z-direction, 2l.

In the present setup, vertically polarized Illuminating ($I_{ill}$) and Reference ($I_{ref}$) beams cross in the particle suspension under an angle $\Theta_S = 6°$. The crossing point is located at the very cell center, taken as origin for the lab frame axes, oriented as shown in Fig. 1b. Typically, cells of rectangular cross-section are employed with length 2l and height 2h, being much larger then cell depth 2d. The height to depth ratio in our case is K = h / d $\geq$ 10. The scattering vector $\mathbf{q} = \mathbf{k}_s - \mathbf{k}_i$ is the momentum transfer on the scattering particle. Its magnitude |**q**| = q = (4π ns / λ$_0$) sin(Θ$_S$/2) [36] depends on the scattering angle, the laser wave length λ and the suspension refractive index n$_S$. Under symmetric illumination, the scattering vector is parallel to the applied electric field E, pointing in positive z-direction. The detection optics ensures that only the light scattered with the wave vector equal to the one of the Reference beam **k**$_f$ = **k**$_{ref}$ is collected from the light shaded detection volume inside the cell. The scattered light and Reference beam mix onto the detector surface. Note that light originating from the Illumination beam and scattered by particles moving in z-direction(upward in Fig. 1a) towards the detector is Doppler shifted with a frequency ω$_D$ = - q·v(x,y,z), where v(x,y,z) is the velocity of the scatterers which may depend on their position. The beat from superimposing Doppler shifted scattered light and unaltered Reference beam light is analysed by a frequency analyser (Ono Sokki DS2000, Compumess, Germany). Typically, 200-2000 individual frames are averaged to yield the power spectrum as a function of linear frequency f = ω / 2π.

We consider the case of scattered light with Gaussian statistics and particles drifting with a constant velocity v$_0$. In addition, we assume the particles to undergo Brownian motion with an effective diffusion coefficient, D$_{eff}$, which may depend on direct and hydrodynamic particle interactions. The super-heterodyne power spectrum C$_{shet}$(q,ω), is the time Fourier transformation of the super-heterodyne mixed-field intensity autocorrelation function, C$_{shet}$(q,τ),:

$$C_{shet}(\mathbf{q},\omega) = \frac{1}{\pi} \int_{-\infty}^{\infty} d\tau \; \cos(\omega\tau) C_{shet}(\mathbf{q},\tau) \quad (1)$$

with circular frequency ω and correlation time τ. Then, the complete power spectrum becomes [31]:

$$\begin{aligned} C_{shet}^0(\mathbf{q},\omega) = & \left[ I_r + \langle I_s(\mathbf{q}) \rangle \right]^2 \delta(\omega) \\ & + \frac{I_r \langle I_s(\mathbf{q}) \rangle}{\pi} \left[ \frac{q^2 D_{eff}}{(\omega + [\omega_B - \omega_D])^2 + (q^2 D_{eff})^2} + \frac{q^2 D_{eff}}{(\omega - [\omega_B - \omega_D])^2 + (q^2 D_{eff})^2} \right] \\ & + \frac{\langle I_s(\mathbf{q}) \rangle^2}{\pi} \frac{2 q^2 D_{eff}}{\omega^2 + (2 q^2 D_{eff})^2} \end{aligned} \quad (2)$$

where I$_r$ is the reference beam intensity, and <I$_s$(q)> is the time-averaged single scattering intensity for the chosen *q*. The subscript *shet* stands to specify the case of super-heterodyning. The equation can also be used for simple heterodyning setting ω$_B$ = 0. The power spectrum contains three terms:

a static background, the super-heterodyne signal and the homodyne signal. Due to the super-heterodyning these terms are well separated in central frequencies. Both the static δ-peak and the homodyne signal are insensitive to the particle drift motion and are centered at zero. The super-heterodyne term consists of two Lorentzians of widths $w = q^2 D_{eff}$, symmetrically shifted away from the origin to $\omega = \pm(\omega_B - \omega_D)$. Due to symmetry, the very same information is contained in each.

## Integral SASH-LDV (ISASH-LDV)

Above, colloids drifting with a constant velocity throughout the detection volume were assumed. However, generally this is not the case in electro-kinetic experiments. Rather one has a constant electro-phoretic velocity $v_{ep} = \mu_{ep} E$ superimposed on the electro-osmotically driven solvent generating position dependent solvent velocities $u_S(x, y)$. These add to the electro-phoretic velocity and for the present geometry, the resulting particle velocity reads:

$$v_P = v_{ep} + u_S(x, y = 0) \tag{3}$$

with particle velocity now becoming a function of the particle position in the cell, which leads to the distribution of velocities in the detection volume.

In the present integral mode, scattered light is collected from the complete cross section at mid cell height. The super-heterodyne signal thus averages over all velocities present in the detection volume. In fact, writing the normalized particle velocity distribution, $p(v) \sim dx/dv$ [30] in terms of the normalized distribution of Doppler frequencies $p(\omega_D)$, the spectrum can be written as convolution integral:

$$C_{shet}(\mathbf{q}, \omega) = \int d\omega_D \, p(\omega_D) C_{shet}^0(\mathbf{q}, \omega) \tag{4}$$

It is worth emphasizing that both homodyne and background terms, shown in eq. 2, stay unaffected by the convolution with the particle velocity distribution. The super-heterodyne part of the spectrum now is a diffusion-broadened distribution of Doppler-frequencies with origin at $\pm \omega_B$. Note further, that for a given field strength, E, the shape of $C_{shet}(\mathbf{q}, \omega)$ is solely determined by the solvent flow profile, i.e. by the electro-phoretic particle and electro-osmotic wall mobility velocity, while the effective particle diffusion coefficient only determines the spectral broadening and the integrated intensity depends on particle concentration and scattering cross section. In a fit of these expressions to experimental data, the five entering parameters are thus mutually independent.

In the present study, the two cases of symmetric and asymmetric electro-osmotic flows are of particular interest. Following [14] we calculated the solvent velocity profiles from a superposition of the flows induced by the electro-osmosis along each of the four walls of a rectangular cell.

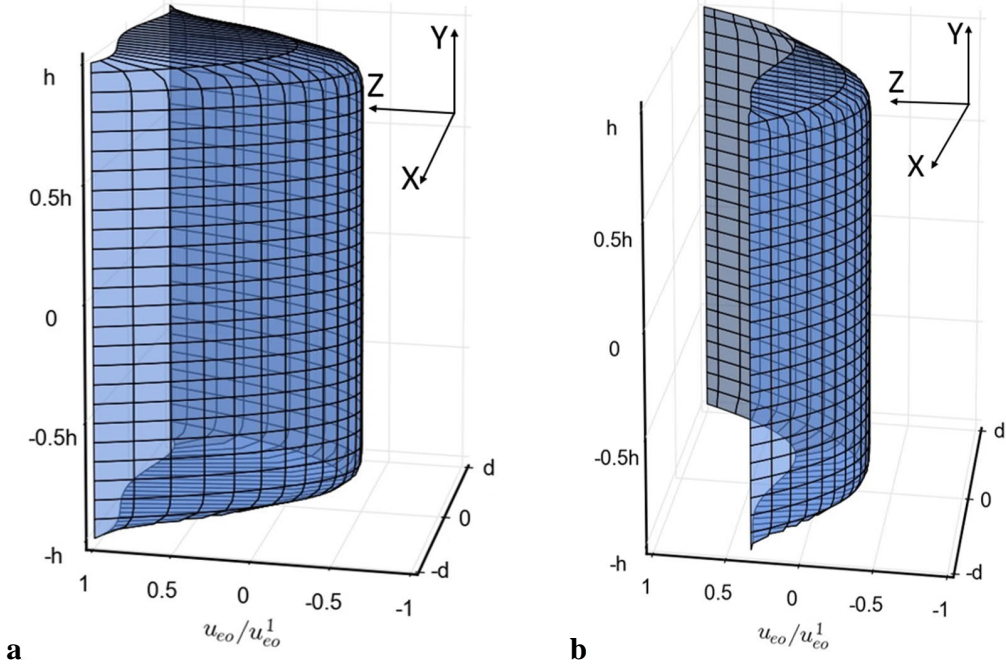

Fig 2: Numerically determined electro-osmotic solvent flow in a cell of K = h / d = 20: a) symmetric, and b) asymmetric flows. Electrode positions are assumed to be as sketched in Fig. 1. Solvent velocities are normalized to the electro-osmotic velocity $u_{eo}^1$ of the symmetric case. In the asymmetric case, $u_{eo}^2 = 1/3 \ u_{eo}^1$. Vertical y-axis: normalized cell height h; x-axis: normalized cell depth d.

In Fig. 2, we show the solvent flows in a real space for a closed rectangular cell with aspect ratio K = h / d = 20. Electrode positions are assumed to be as in Fig. 1. Thus the electro-osmotic velocity $u_{eo}$ at the negatively charged surfaces, points towards the left, i.e. in positive direction. Due to the solvent incompressibility, the net flow is zero [11] and the electro-osmotic flow at the walls is balanced by the solvent back-flow at the cell centre. In Fig. 2a we show the case of equal electro-osmotic velocity $u_{eo}^1$ at both sidewalls. The resulting flows resemble parabolas with corrections for the presence of the upper and lower walls. For symmetric flows at mid-cell height, an analytic approximation is available [37]:

$$v_P(x, y=0, z) = \mu_{ep}E + \mu_{eo}E\left[1 - 3\left(\frac{1 - \frac{x^2}{d^2}}{2 - \frac{384}{\pi^5 K}}\right)\right] \tag{5}$$

In Fig. 2b we show the case of the front wall having a lower velocity $u_{eo}^2 = 1/3\ u_{eo}^1$. The solvent flow at mid-cell height, i. e. in the scattering plane, still resembles a parabola. However, the maximum velocity is observed closer to the low mobility wall. This can also be seen in Fig. 3, where we plot the particle velocity profile (red) for the asymmetric case of Fig. 2b. As compared to the solvent flow profile (yellow) the former curve appears shifted by the particle electro-phoretic velocity. In order to avoid cancellation of wall velocities, $v_{ep} = -2/9\,\mu_{eo}^1$ was chosen to be negative and somewhat smaller than $u_{eo}^2$. In blue we show the corresponding normalized velocity distribution, p(v), which is peaked at the maximum particle velocity. Note that unlike in the symmetric case the maximal velocity is not at the cell centre. p(v) features two steps located at $u_{eo}^1 + v_{ep} = E(\mu_{eo}^1 - v_{ep})$ and $u_{eo}^2 + v_{ep} = E(\mu_{eo}^2 - v_{ep})$. The step in p(v) corresponding to the lower $u_{eo}^2$ is characteristic for the asymmetric case and not seen in the symmetric case.

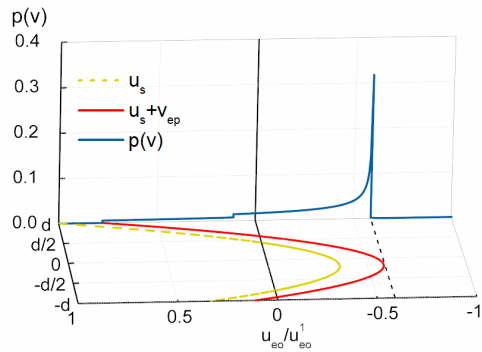

Fig. 3: Mid-cell height solvent (dashed yellow) and particle velocity (solid red) profile and resulting normalized particle velocity distribution (solid blue) for the asymmetric case shown in Fig. 2b and $v_{ep} = -2/9\,\mu_{eo}^1$.

Inserting the calculated mid-cell height particle velocity distributions in Eqn. 4 yields the super-heterodyne power spectrum. A complete shet-spectrum is shown in Fig. 4 for the case of asymmetric electro-osmotic flow discussed above. It contains a δ-function at the origin corresponding to the static background in Eqn. 2. The broad homodyne term with width $w_{homo}$ =

$2q^2D_{eff}$ also centred at the origin. The super-heterodyne parts are shifted to $\omega = \pm\omega_B$, here we chose $\omega_B$ = 2 kHz for display reasons. They reflect existing velocity distribution in the cell and are diffusion broadened with $w_{shet} = q^2D_{eff}$. As usual, the different orientation of the features allows discriminating the sign of wall and particle surface charge. In the present case one infers a negative electro-phoretic velocity. Since the information contained in each is identical, we will only use the spectral region around positive $\omega_B$.

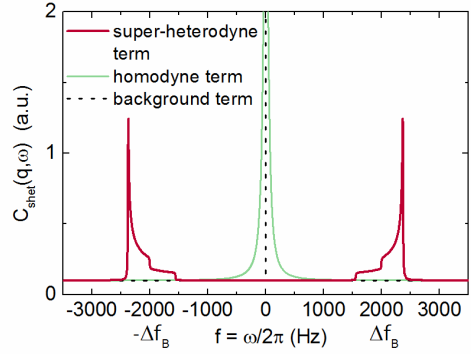

Fig. 4: Complete super-heterodyne spectrum for the case of asymmetric electro-osmotic solvent flow. Parameters used for calculation: $u_{eo}^1$ = 135 µms$^{-1}$, $u_{eo}^2$ = 45 µms$^{-1}$; $v_{ep}$ = 15 µm$^{-1}$, $D_{eff}$ = 2 10$^{-12}$ m$^2$s$^{-1}$, K = 20 and $\omega_B/2\pi$ = 2 kHz.

Evaluation of measured spectra starts with subtracting the frequency independent noise background and isolating the frequency range of interest, centered about +$\omega_B$. We then make an estimate of electro-osmotic velocities, $u_{eo}^1$ and $u_{eo}^2$, and electro-phoretic velocity, $v_{ep}$, from the location of the frequencies of the maximum and the steps. Further evaluation employs a self-written Python script. First, the particle flow profile is calculated from the estimated velocities. Here, the cell aspect *K*, y-position of the measurement plane and the strength of applied electric field *E* are constants, known from the experiment. Next, the particle velocity distribution is obtained from differentiating the flow profile p(v) ~ dx / dv, where x is the position of the particle in the scattering plane. From p(v) the distribution of Doppler frequencies, p($\omega_D$), is calculated using $\omega_D$= q·v. A super-heterodyne Lorentzian is calculated from Eq. 2 with particle diffusion coefficient $D_{eff}$ as input parameter and experimental constants q and $\omega_B$. This Lorentzian is convoluted with the p($\omega_D$) according to Eq. 4. Finally, the integrated intensity is adjusted with A, the signal amplitude, being again an independent parameter depending on the particle concentration and scattering cross section, which remains constant for each measurement series. The calculated theoretical $C_{shet}(q,\omega)$ signal is then used in the least square fit of the experimentally measured super-heterodyne signal over the frequency range of interest, i.e. centred around $\omega_B$. The fit is performed with Levenberg-Marquardt algorithm, implemented in the SciPy library [38]. We stress again that due to the functional form of $C_{shet}(q,\omega)$, the four physically relevant parameters, $u_{eo}^1$, $u_{eo}^2$, $v_{ep}$ and $D_{eff}$ are mutually independent and define the signal width, location of the steps, position of the center of the signal and diffusive broadening of the spectrum. Electro-kinetic velocities are then plotted in dependence on electric field strengths and a linear fit yields the corresponding mobilities as slopes.

**Sample preparation and characterization**

Particles for measuring electro-phoretic mobilities were commercial polystyrene latex spheres (lab code PS301, manufacturer batch #10-66-58. IDC, Portland, USA) nominal diameter $2a = 301$ nm; nominal polyispersity index $Pi = 0.08$ both as given by the manufacturer; diameter from dynamic light scattering (DLS) $2a = 322 \pm 2.4$; diameter from static light scattering (SLS) $2a = 310 \pm 4.0$ nm [39]. Particles are stabilized by sulphate surface groups of titrated charge number $N = (2.3 \pm 0.2) \times 10^4$ [14]. The stock suspension was diluted with milli-Q grade water to the desired number densities and stored prior to use on mixed bed ion-exchange resin (Amberlite K306, Carl Roth GmbH + Co. KG, Karlsruhe, Germany) in gas tight screw cap vessels.

**Sample conditioning**

All the experiments were conducted in a conditioning circuit to ensure stable and well-controlled experimental conditions. The circuit consists of an electro-kinetic cell, a sample reservoir under inert gas atmosphere (to add water, stock-suspension or electrolyte solution), an ion exchanger column (Amberjet, Carl Roth GmbH + Co. KG, Karlsruhe, Germany) with bypass for the deionization of the colloidal suspension and a conductometric experiment (electrodes LR325/01 bridge LF340i, WTW, Germany) for monitoring particle concentration in the deionized state and electrolyte concentration after fixing the particle concentration [40]. Any additional number of experimental cells can be connected, e.g. one for static light scattering to monitor the suspension structure. Experimental parameters can be adjusted to desired values within short times and the suspensions are thoroughly homogenized. For the present experiments, the particle number density was adjusted to $n = 4.5 \; 10^{15}$ m$^{-3}$ to obtain multiple scattering free signals. Our circuit conditioning technique allows measurements of colloidal dispersions at reproducibly adjusted low salinities, without a need of a buffer, which introduces extra ions in the system. To this end, the suspension is first thoroughly deionized to minimum conductivity, then the ion exchanger is bypassed and the Ar-atmosphere of the reservoir is exchanged by ambient air and the suspension cycled until the $CO_2$ dissolution and dissociation reactions have equilibrated [41]. At a constant ambient temperature of 23.5° C, the electrolyte concentration is thus maintained at $c = 5 \cdot 10^{-6}$ mol l$^{-1}$ for all experiments. This in additions ensures comparability to the micro-fluidic experiments performed at the same temperature and in contact with ambient air.

**Electro-kinetic cell**

All electro-kinetic experiments were conducted in a custom build flow-through cell made of Poly-Methyl-Metacrylate (PMMA) as shown in Fig. 5. The two slide panels are equipped with sealing rings and are tightly screwed to the main body. Each slide panels fixes one charged wall specimen (standard microscopy slide) placed in the main body frame. Use of gas tight o-rings is essential to avoid air bubbles and to seal the cell from atmospheric carbon dioxide, when working at completely deionized conditions. The two electrode chambers are fitted to the left and right end of the main body as can be seen in the top left. Platinized platinum electrodes are inserted into the electrode holders of the electrode chamber. The effective electrode distance of $l = 11.2$ cm$^{-1}$ was calibrated from conductivity measurements on standard electrolyte solutions. Each chamber has two additional bores, one for filling and one for an optional thermocouple. For filling and adjustment of experimental parameters, the cell is connected at each electrode chamber via Teflon® tubings to a peristaltically driven conditioning circuit [42].

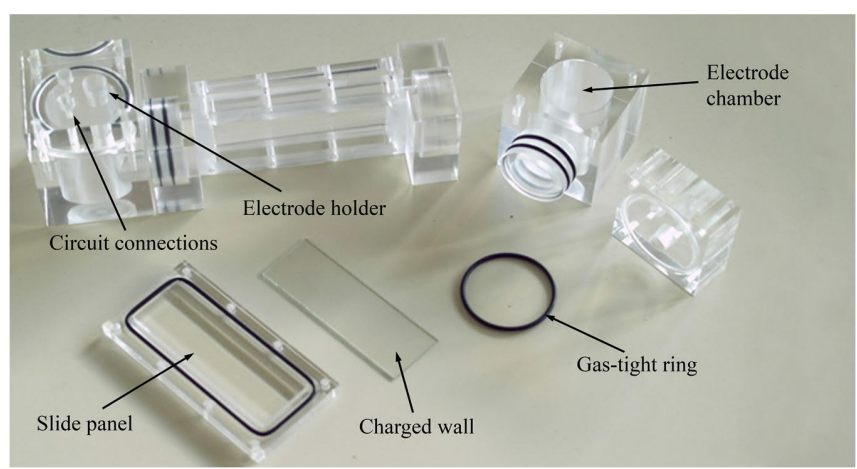

Fig. 5: Electro-kinetic cell with exchangeable sidewalls. The charged wall specimens are placed into the PMMA main body frame and fixed by the slide panels. O-rings are used to render the cell

interior liquid- and gas-tight. Electrodes are inserted into the electrode chambers which in turn are inserted into the main body frame. The sample conditioning circuit (not shown) is connected at each electrode chamber at the screw docks denoted circuit connections.

Symmetric electro-osmotic flows are achieved using two identical specimen slides. Standard microscopy glass slides (75 x 25 x 1 mm, soda lime glass of hydrolytic class 3 by VWR International, Germany) were sonicated prior to use for 60 min in 2% alkaline detergent water solution (Hellmanex III, Hellma Analytics) at 35 $^0$C and rinsed with doubly distilled water several times.

Asymmetric flows are achieved using sidewalls which either were made of different materials or were treated differently. To obtain moderately low charged specimens, PMMA sheets of thickness 1 mm were cut to size of 75x25 mm$^2$ corresponding to a standard microscopy slide size. Cleaning and washing was done by the same procedure as for glass slides. Very low charge specimens were prepared by coating with N,N-Dimethyl-N-octadecyl-3-aminopropyltrimethoxysilyl chloride (DMOAP). Since the coating ages over time, fresh specimens were prepared for each experiment and instantaneously used. Pre-cleaned glass slides were immersed into 1% DMOAP aqueous solution for 2 minutes, removed, rinsed with doubly distilled water and dried in oven under nitrogen atmosphere at 70°C for 60 min.

Interestingly, there is no perfect standard for the electro-osmotic mobility. Even such well known substrates as glass, quartz or PMMA are reported to have different mobilities, dependent on the exact composition, history of the material, presence of impurities or pH of the system [43, 44]. We therefore also try an *internal* standard, which provides equal mobilities for the suspension particles and one of the sidewalls. As compared to using just the particles as calibration standard, our approach has the advantage of eliminating effects of electro-static particle interactions, of charge renormalization and charge regulation (which differ for spheres and charged walls [45, 46]), and in particular the experimentally well known density dependence of particle mobilities [28]. The envisioned situation is sketched in Fig. 6a.

To prepare the DMOAP/PS301 coated slides, pre-cleaned glass slides were immersed into 2% DMOAP solution, removed and dried in the oven under nitrogen atmosphere at 70 °C for 60 min. The slides were then rinsed with double distilled water, brought into contact with a suspension of PS301 (n = 1 10$^{16}$ m$^{-3}$) and left over night. The next day, the PS301-coated slides were removed and rinsed with distilled water. The positively charged quaternary nitrogen atom in DMOAP molecules electrostatically attracts the negatively charged polystyrene particles, thus acting as a primary physisorbing agent. Subsequently, van der Waals binding of a monolayer of PS301 occurs. A typical result of this coating protocol is shown in Fig. 6b.

Theoretically, electro-osmotic performance of such a coating by known particles could depend on topological roughness as well as electro-kinetic roughness, i.e. the spacing between particles, acting as differently conductive areas [47, 48]. For sufficiently close packed monolayer and at low salt conditions, however, we expect the double layers of the coating particles to overlap and dominate completely over the DMOAP double layer. The resulting zeta-potential and electro-osmotic mobility should hence equal the surface potential of suspended PS301 and their electro-phoretic mobility, respectively. We will check this approach for a useful internal standard in the results section.

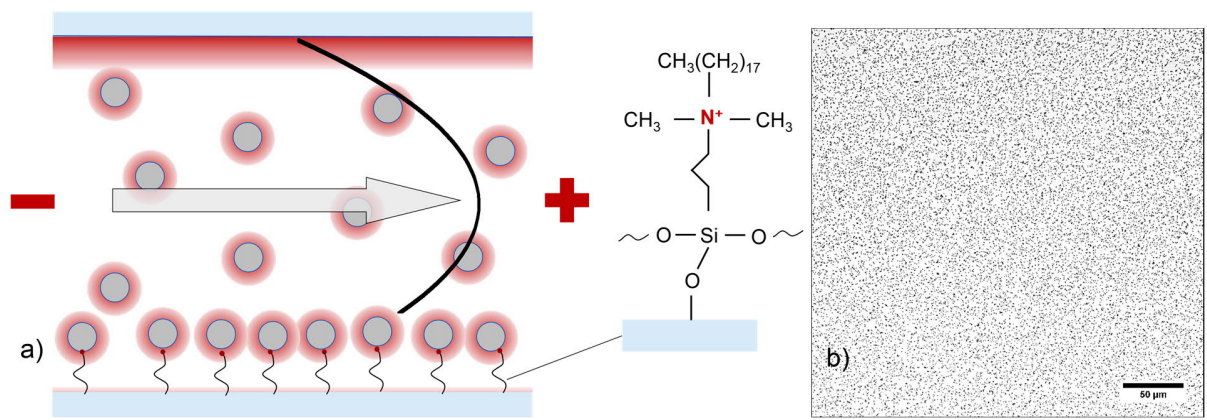

Fig. 6: Scheme for internal electro-phoretic-electro-osmotic standard: a) schematic sketch of the electro-kinetic cell with two different specimens mounted (not to scale). Location of anode and cathode are indicated by the plus and minus sign, the electro-phoretic flow direction is indicated by the arrow and the resulting flow profile by the solid line. Note that diffuse parts of electric double layers are drawn out to distances, where $e\varphi = k_B T$ with potential values qualitatively indicated by colour saturation. The top wall of uncoated glass is highly charged, the DMOAP-coated bottom wall only weakly. Charge there is provided by the monolayer of PS301. Note that at low screening and sufficiently large areal coverage, lateral double layer overlap may occur. At the very right, a grafted DMOAP unit is sketched b) Phase contrast image of a PS301-coated wall, showing the attached polystyrene particles as black dots. The areal coverage is about 17%. Scale bar 50µm.

## Microfluidic experiments using osmotic pumps

Using the newly designed cell with internal standard we can measure the mobilities of charged spheres and one unknown substrate using the other as internal standard. In the worked example discussed below, these values are employed to interpret a micro-fluidic pump transporting charged-stabilized tracers by electro-osmotic flow. The specific pump investigated utilizes a single ion exchange resin bead (IEX) placed on a negatively charged substrate. The experimental set-up and the involved transport mechanism have recently been described in detail [25]. The IEX exchanges residual cationic impurities for protons and thus creates a radially symmetric pH-gradient [49]. Differing mobilities of involved impurity cations and protons give rise to a radially decaying diffusio-electric field. This in turn drives a convergent electro-osmotic flow along the substrate carrying along the colloidal tracers, which themselves are electro-phoretically driven in outward direction. Theoretical modelling of tracer transport and accumulation therefore requires the knowledge of the electro-osmotic mobility of the substrate and the tracers. Using the conditioning circuit we can conveniently reproduce the experimental conditions of the pumping experiments and provide the required electro-osmotic mobility.

To be specific, the pumping experiment employs commercial spherical IEX particles of diameter of 45±1 µm (CGC50×8, Purolite Ltd, UK; lab code IEX45) for field generation. These are made of sulfonated cross-linked poly(styrene-divinylbenzene) copolymers (cross-linking degree: 8%) with proton as exchangeable cation. Tracers were commercial carboxylate stabilized polystyrene spheres with diameter 2a = 15.2±0.1 µm (MicroParticles GmbH, Germany batch No. PS/Q-F-L1488; lab code PS15). The initial 10% w/v dispersion was diluted with doubly distilled water and then thoroughly deionized in contact with ion-exchange resin (Amberlite K306, Roth GmbH, Germany). Under deionized but $CO_2$ saturated conditions, the electro-phoretic mobility of this rapidly settling species was found to be $\mu_{ep}$ = 2.5±0.2 $m^2$ $V^{-1}$ $s^{-1}$ [50].

Pumping was performed in a closed cell custom made from PMMA rings with inner diameter d = 20 mm and height H = 1 mm. The rings were glued to the different substrates with the hydrolytically inert epoxy glue (UHU Plus Sofortfest, UHU GmbH, Germany). Glass, PMMA and DMOAP-coated glass microscopy slides were used directly after determining their electro-osmotic mobilities in the ISASH-LDV experiments. A single bead of IEX45 was glued to the substrate of each cell with epoxy glue and left to cure for the 24 hours under dust-free conditions. We then added 0.4 mL of diluted dispersion of PS15 and quickly sealed the cell with an upper slide to avoid contamination with dust during the pumping experiments.

The cells were mounted on the stage of an inverted scientific microscope (DMI4000 B, Leica, Wetzlar, Germany) equipped with a standard video camera and observed in bright field with 5x or 10x magnification objectives. Videos were collected at a frame rate of 1 Hz and processed with a self-written Python script. From the positions of tracer particles in subsequent frames the velocities were calculated and for different type of substrate monitored as a function of distance from the ion exchange resin centre. From the theoretical analysis of the electro-osmotic pump [25], we would expect the integrated solvent flux to be a linear function of the electro-osmotic substrate velocity.

## Results

We first address symmetric electro-osmotic flows. In Fig. 7a we show the heterodyne part of the spectrum which appears as a diffusion broadened velocity distribution of the symmetric solvent flow residing on a frequency independent white noise background at a level of $1 \; 10^{-5}$ V Hz$^{-1/2}$. Measurements are repeated at different field strengths. The corresponding data are displayed in Fig. 7b. To check absence of flow distortion by shear thinning at large electro-osmotic flows, of field induced density fluctuations and of other non-linearities, we perform field strength scaling. We subtract the Bragg shift frequency and divide the spectral range by E, while multiplying the signal amplitude by E [31]. In Fig. 7c, all spectra are seen to neatly collapse on a single curve, thus demonstrating the absence of unwanted flow or spectral distortions. Each data set was fitted as described above and the least square fits are shown in Fig. 7b as dark solid lines. As an example, the fit for E = 16.8 V cm$^{-1}$ returned $u_{eo}$ = 73.2 ± 3 µm s$^{-1}$, $v_{ep}$ = 55.2 ± 7 µm s$^{-1}$ and $D_{eff}$ = 3.4 $10^{-12}$ m$^2$ s$^{-1}$.

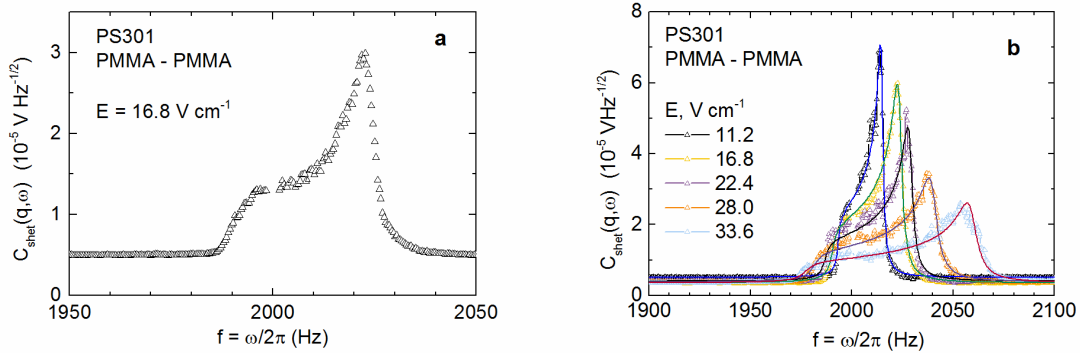

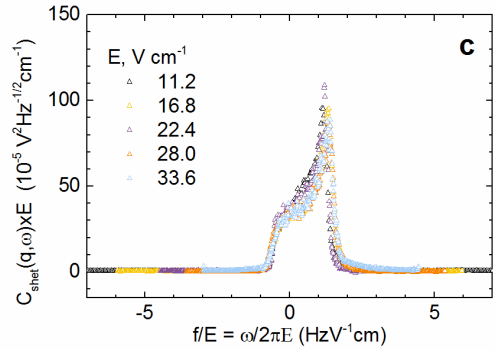

Fig. 7: Electrokinetic results of the symmetric flow case. (a) Typical spectrum as obtained at E = 16.8 V cm$^{-1}$ of PS301 at n = 4.5 10$^{15}$ m$^{-3}$ in a cell with two PMMA walls. (b) Spectra obtained for this situation for different field strengths as indicated. Fits of theoretical expressions to the data are shown as solid lines. (c) Background corrected and field strength scaled spectra collapse to a single curve testifying the absence field or flow induced non-linearities

A compilation of the electro-phoretic velocities of PS301 measured in cells with pairs of identical walls of different materials in the field strength range of E = (10-35) Vcm$^{-1}$ is displayed in Fig.8a. A least square fit (dashed line) returns $\mu_{ep}$ = 3.46 ± 0.26 10$^{-8}$ m$^2$ V$^{-1}$ s$^{-1}$, i.e with a relative uncertainty on the order of 5%.

The results for the electro-osmotic velocities of the different materials are displayed in Fig. 8b. Here, different electro-osmotic mobilities are obtained from the least square fits of lines through the origin. We obtain for glass, PMMA and DMOAP: $\mu_{eo}$ = 11.0 ± 0.6 10$^{-8}$ m$^2$ V$^{-1}$ s$^{-1}$, $\mu_{eo}$ = 4.75 ± 0.35 10$^{-8}$ m$^2$ V$^{-1}$ s$^{-1}$, and $\mu_{eo}$ = 1.55 ± 0.20 10$^{-8}$ m$^2$ V$^{-1}$ s$^{-1}$, respectively. As expected, the electro-phoretic mobility of colloidal particles is independent of the wall materials of the electro-kinetic cell, while the electro-osmotic mobilities differ considerably. It may be worth mentioning that the electric field scaled data, e.g. Fig. 7c, allow an alternative way of determining the mobilities with comparable uncertainty level. Here, one averages the normalized spectra and then proceed with a single standard signal fitting procedure.

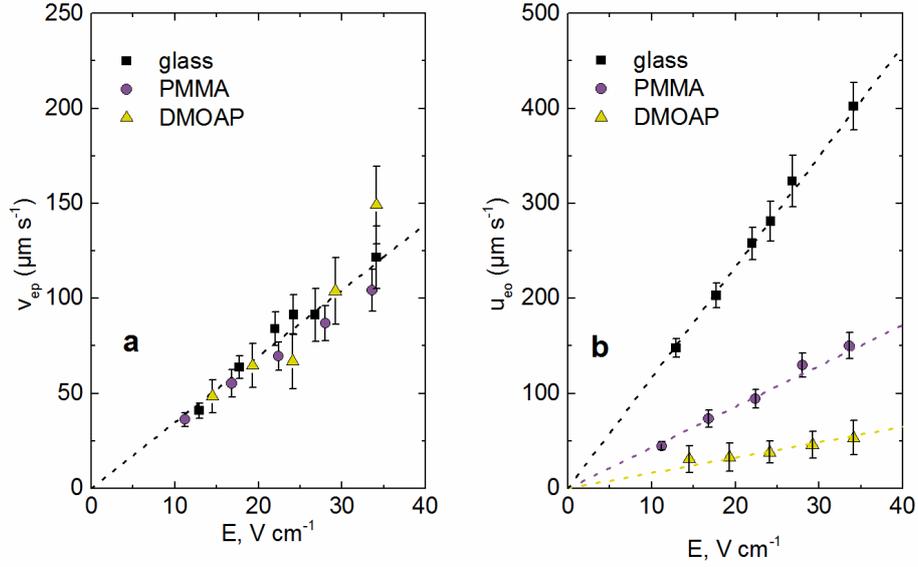

Fig. 8: Field dependence of the electro-kinetic velocities. (a) electro-phoretic velocities of PS301 in cells with two identical sidewalls made from different materials as indicated. Data coincide on a straight line through the origin. A least square fit (dashed line) returns $\mu_{ep} = 3.46 \pm 0.30 \; 10^{-8}$ m$^2$ V$^{-1}$ s$^{-1}$. (b) Field dependence of the electro-osmotic velocities for different sidewalls as indicated.

We next turn to asymmetric flows. Fig. 9 compares the heterodyne spectra of PS301 of a) symmetric and b) asymmetric flow under otherwise identical conditions. Note the qualitatively different appearance of the spectra, which is highlighted in Fig. 9c after background correction. The spectrum of the symmetric flow situation appears somewhat stretched due to the larger sum of electro-osmotic velocities. The integrated intensity coincides within experimental uncertainty $A_{glass-PS} = 7.2 \pm 0.3 \; 10^{-4}$ V Hz$^{1/2}$ and $A_{glass-glass} = 6.7 \pm 0.3 \; 10^{-4}$ V Hz$^{1/2}$. This again testifies the absence of non-linearities and significant differences in sample preparation. Moreover, for both situations the fit returns: $u_{eo}^1 = 400 \pm 25$ µm s$^{-1}$, $v_{ep} = 121 \pm 16$ µm s$^{-1}$ and $D_{eff} = 1.1 \; 10^{-11}$ m$^2$ s$^{-1}$. In addition, we have $u_{eo}^2 = 119 \pm 17$ µm s$^{-1}$ for the data in Fig. 9b.

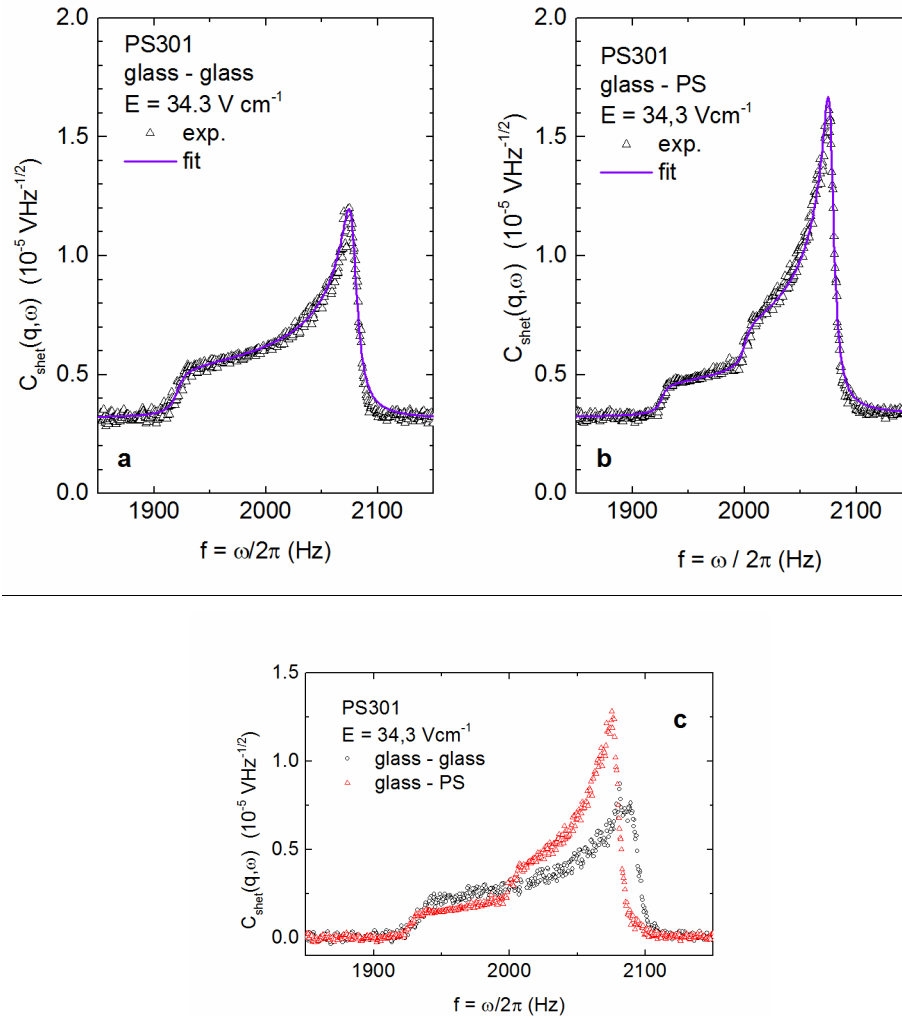

Fig. 9: Characteristic spectra for symmetric and asymmetric electro-kinetic flow of PS301 at n = 4.5 $10^{15}$ m$^{-3}$ and the same electric field. (a) Measured in a cell with two glass walls. (b) Measured in a cell with one glass wall and one DMOAP/PS301-coated wall. The step corresponding to the particle velocity at the coated wall is clearly visible. It appears exactly at the Bragg shift frequency $f_B$ = 2000 kHz, since the electro-phoretic and the electro-osmotic velocities cancel each other close to the coated wall. (c) Comparison of the background corrected spectra of (a) and b). The overall integrated intensity stays constant but is considerably redistributed as the flow is switched from symmetric to asymmetric.

A series of spectra taken at different field strengths is displayed in Fig. 10a. As the field strength increases, the spectra stretch out and slightly shift to the right. All spectra are perfectly described by the corresponding least square fits. Fig. 10b compares the fit results obtained for the glass-glass case to those obtained for the glass-PS case. All derived electro-kinetic velocities depend linearly on the applied field strength indicating a stable coating. The field strength dependent electro-osmotic velocities of the glass walls coincide neatly ($\mu_{eo}$ = 1.10 ± 0.06 $10^{-8}$ m$^2$ V$^{-1}$ s$^{-1}$), as do the electro-phoretic velocities of PS301 in both cases ($\mu_{ep}$ = 3.64 ± 0.11 $10^{-8}$ m$^2$ V$^{-1}$ s$^{-1}$). Most interestingly, the latter coincide quantitatively with the field dependent electro-osmotic velocities of the PS-coated wall ($\mu_{eo,PS}$ = 3.61 ± 0.17 $10^{-8}$ m$^2$ V$^{-1}$ s$^{-1}$).

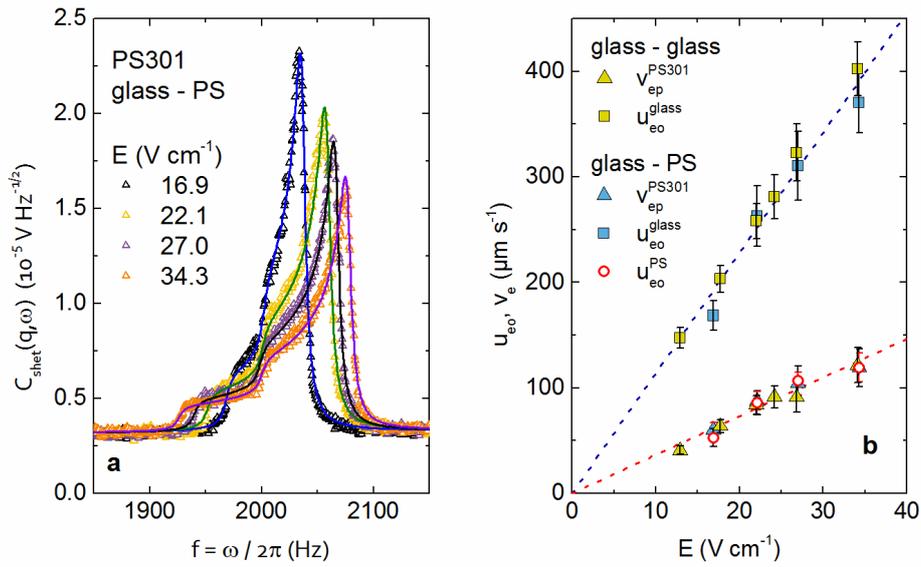

Fig.10: Field dependent electro-kinetic spectra (a) and velocities (b) obtained for PS301 at n = 4.5 $10^{15}$ m$^{-3}$ for symmetric (glass-glass) and asymmetric (glass-PS) flows. In all cases we observe strictly linear field dependences of the velocities. From the slopes of linear least square fits, shown as dashed lines, we obtain $\mu_{eo, glass}$ = 1.10 ± 0.06 $10^{-7}$ m$^2$ V$^{-1}$ s$^{-1}$, $\mu_{eo,PS}$ = 3.61 ± 0.17 $10^{-8}$ m$^2$ V$^{-1}$ s$^{-1}$ and $\mu_{ep}$ = 3.64 ± 0.11 $10^{-8}$ m$^2$ V$^{-1}$ s$^{-1}$

## Discussion

The present investigation focused on simultaneous determination of electro-osmotic and electro-phoretic mobilities exploiting the possibility of integral measurements in a small angle super-heterodyne LDV. We designed a flow-through electro-kinetic cell featuring exchangeable sidewalls to mount standard microscopy slides of different materials or with different coatings and built-in connections to a conditioning circuit to adjust the tracer and electrolyte concentrations under conductometric control. We performed experiments for charged colloidal spheres and different substrates in dependence on applied field strength to demonstrate the performance of our instrument and the newly designed cell.

We start our discussion with a few more general remarks on ISASH-LDV, which has been introduced in more detail in [16]. In the *integral* configuration, the Doppler-shifts $\omega_D = q \cdot v_P$ are collected from the complete cross-section of this electro-kinetic cell at mid-cell height. Measurements thus cover the complete velocity profile and the obtained spectra correspond to velocity distributions. Numerical calculations of Poiseuille type velocity profiles and corresponding normalized particle velocity distributions from published theoretical expressions were implemented for symmetric and asymmetric flows in the electro-kinetic cell. These were used in combination with the theoretical expressions, Eqs. 2 and 4, for fitting the super-heterodyne part of the measured spectra. Each fit returns independent values of all involved electro-kinetic velocities and an effective diffusion coefficient. A second linear fit of the results returns the desired mobilities. The simultaneous measurements of both relevant mobilities in a single experiment eliminate the need for sophisticated additional measurements on which corrections for electro-osmotic flows are to be based [20, 21]. We demonstrated the excellent performance of this approach for both symmetric and asymmetric flows.

Concerning the precision of our measurements, we note that in the investigated range of field strengths, the DMOAP coated substrates show rather narrow signals and therefore larger fit uncertainties, due to the decreased amount of data points per signal and stronger effect of diffusive broadening onto the spectrum. This is reflected in the relative size of the error bars in Fig. 8a and b. Nevertheless, electro-phoretic mobilities can be determined with a relative statistical uncertainty on the order of 5%. Systematic uncertainties in preparation do typically not exceed 1% for the particle concentration and 2% for the electrolyte concentration. Electro-osmotic mobilities are determined with similar statistical precision but without systematic error in tracer concentration. In the case of small electro-osmotic flows the total uncertainty of electro-osmotic mobilities in the presently investigated field strength range was about 10%. However, this can be conveniently compensated by employing larger field strengths leading to a larger Doppler-frequencies and a larger stretching of the spectral shape as compared to diffusive broadening. The obtained precision therefore competes well with that of commercially available devices.

Use of the integral configuration, however, offers also clear advantages over most conventional approaches. First, the same precision is obtained at much less time. Both micro electro-phoresis and Phase Analysis Light Scattering (PALS) [29] are restricted to point-by-point real space measurements of the flow profiles. Second, ISASH-LDV allows a facile check procedure to detect factors influencing accuracy, i.e. any deviations from the theoretically expected flow behaviour. These include spatially dependent non-Newtonean behaviour, non-Poiseuille type flows, non-linear electro-kinetics, and/or medium scale longitudinal or transversal density fluctuations under electro-kinetic motion or shear. Non-Poisseuille flows, observed e.g. in shear thinning samples, significantly distort the spectral shape in a field dependent way and field scaling (Fig. 7c) is lost. Here, the osmotic mobilities stay well defined. More complicated is the case of colloidal crystals adhering to the cell wall despite fully functioning electro-osmotic flow [34]. Then the electro-osmotic velocity cannot be determined at all. Still the electro-phoretic particle velocity can be inferred from a double average over the complete cell depth and cell height, $x$ and $y$ [31]. Medium scale density fluctuations involving many particles may result from hydrodynamically induced velocity fluctuations under electro-phoretic particle motion in a homogeneous applied electric field [51]. Similarly, lateral density fluctuations may occur induced by the electro-osmotic shear field [52] In our small angle configuration, both in principle may interfere with sample homogeneity and again would lead to a loss of scaling. Also non-linear electro-kinetic effects, e.g. a field strength dependent electro-phoretic or -osmotic mobility [53], destroys the field scaling. For a single experimental run at a given field strength about 10 min are needed to reach the statistical accuracy discussed above. For a quick field scaling check, however, the measurement time can be significantly reduced and the sample thus conveniently controlled for expected (theory compliant) performance. We performed such checks of some ten min total duration after each sample preparation. In all cases perfect field scaling was observed. This possibility offers a significant advantage over many commercial devices where deviations between theoretically expected and experimentally realized flow profiles are much harder to check.

In the present ISASH-LDV experiments the focus was put on electro-osmosis along different substrates, keeping the colloidal (tracer) system constant. The custom made cell with exchangeable side-walls allowed for a flexible and quick characterization of different negatively charged wall materials. The obtained wall mobilities can be converted to zeta-potentials using standard electro-kinetic theory [1] assuming constant charge boundary conditions and accounting for the tracer particle counter-ion contribution to the total electrolyte content. For the investigated substrates we obtain: $\zeta_{glass}$ = -138.2 ± 8.0 mV, $\zeta_{PMMA}$ = -58.7 ± 4.4 mV and $\zeta_{DMOAP}$ = -19.5 ± 2.5 mV. The zeta-potential of glass slides is in accordance with the values reported for the glass in contact with aqueous solutions of low ionic strength [54]. As pointed out by Barz et al. [20], there is a huge discrepancy between the results obtained by different research groups concerning the PMMA zeta potentials. The potential of PMMA obtained in our experiment is close to the values reported by these authors, when scaled according to their suggested procedure. No literature values had been available for the DMOAP-coated walls but it has frequently been noted, that a very low zeta-potential is obtained as the isoelectric point is approached by suitable methods [3]. For DMOAP-coating, the resulting wall charge depends crucially on the coating protocol. We found that low negative potentials on the order of 1 $k_BT$ could be realized for low DMOAP concentrations and short contact times. Larger concentrations and longer contact times led to positively charged walls, which cannot be measured with negatively charged tracers due to strong particle adsorption effects.

We next turn to some aspects of instrumental flexibility. We here quantified electro-kinetic mobilities of water-based systems. However, ISASH-LDV, is a very general method. There is no principle restriction to the choice of particles, solvents and/or wall substrates as long as system stability (coagulation, adsorption at the wall) is respected. ISASH-LDV could thus be applied for electro-kinetic experiments in non-polar media, which are important for many industrial applications. From the scientific point, these provide interesting challenges concerning the field dependence of the electro-phoretic mobilities [55;56]. Here, our method could contribute to facile in-detail studies on diffuse layer deformation or counter-ion stripping, as well as charge regulation phenomena in organic media [57;58;59]. Only for AC electro-phoresis experiments, the use of ISASH-LDV appears to be limited. There, the signal shape is given as a multiplication of the Fourier transform of the field switching frequency and waveform with the Doppler spectrum. The latter is a continuous signal, while the former shows a series of peaks at multiples of the driving frequency. At AC frequencies much larger than Bragg shift frequency no super-heterodyne spectrum will appear at all. At AC frequencies lower than the one for electro-kinetic Doppler shift, both spectral forms combine resulting in "combed" signal shape of which the envelope still contains some information on the average particle velocity [60]. If electro-osmosis is supressed by the field frequency [29], it can be used to infer electro-phoretic mobilities [61].

By design, our cell allows mounting of different substrates leading to asymmetric electro-osmotic flows. We demonstrated that this can be conveniently respected by implementing the corresponding flow profiles in the evaluation procedure. Moreover, this feature opened the possibility of realizing an internal mobility standard: deliberately introducing strong particle adsorption we covered DMOAP treated walls with PS301. The protocol given in the experimental section yields a high areal coverage with a reproducibility of better than 5% for subsequent coating runs. The coating was sufficiently dense to produce the same substrate mobility as the bulk tracer mobility under the prevailing experimental conditions of comparable inter-particle distances in the bulk and on the surface. This is an inevitable condition for an internal mobility standard. The PS301 coated wall electro-osmotic mobilities were perfectly reproducible and quantitatively coincided in each case with the PS301 particle mobility. Such a potential switching appears suitable for many materials and opens a general way of producing internal standards for electro-phoretic and electro-osmotic mobility measurements in material science.

An internal standard as well as the possibility to compare electro-osmotic and electro-phoretic velocities for the same material, appears to be very interesting also for some long-standing fundamental issues in electro-kinetics. One of these is the experimentally well documented density dependence of the bulk electro-phoretic mobility [28;31;33;58;61]. This quantity shows a maximum at intermediate concentrations. There measured mobilities are typically in excellent agreement with the expectations of the standard electro-kinetic model (SEM) [1]. Towards large particle concentrations, the mobility decreases due to the increase in electrolyte concentration as provided by the particle counter-ions [62]. The decrease towards low particle concentration [33] and the extremely low mobilities of isolated particles [22; 23] are neither anticipated nor explained by SEM or its recent extensions [63]. This unexpected density dependence has tentatively been attributed to the loss of double layer overlap upon dilution [61], but also charge renormalization and/or charge regulation effects may play an important role [64]. Using samples of different particle concentration and substrates with different tracer coverage, a comparison of the density dependence of electro-phoretic and electro-osmotic mobilities obtained from ISASH-LDV in our new cell under identical experimental conditions could further test existing ideas.

Realized as table-top experiments, ISASH-LDV (like PALS [29]) may in principle accommodate any custom made cell designed for specific interests. This releases the restriction to electro-kinetic experiments. In particular, its application to flow measurements under hydrostatic pressure difference [65] is straightforward. There, also flows deviating from Hagen–Poiseuille [66] can be measured and even quantitatively evaluated, if the functional form for the flow profile is available.

In electro-kinetics, the case of very small electro-phoretic mobilities is probably still best addressed in the Uzgiris-type narrow gap electrode cell without interference of electro-osmosis [10, 11], as demonstrated using PALS. In combination with micro-electro-phoresis, such cells are also well suited for strongly sedimenting particles, which cannot be studied well by scattering methods [67]. For very small electro-osmotic velocities, the here presented cell appears to be of great advantage as it allows an internal calibration standard and may use tracers of large electro-phoretic mobility. Cells connecting to a well controlled external conditioning are also easily implemented. We note that our previous experiments at thoroughly deionized conditions have been performed using this approach in commercial optical flat cells first used in the seminal MARK II instrument of Rank. Bros. [30, 31, 32, 33, 34]. These are also available with very thin cross section, making them ideally suited for ISASH-LDV on multiply scattering samples [16].

In principle, the integration even of actual microfluidic devices is possible given these have parallel, flat and transparent front and back walls. To study non-transparent wall materials the design of Burns and others [13, 14, 15] could be used. Such studies appear to be very useful for applications like ellipsometry, ESCA, or contact angle and other wetting-type measurements. The design of our electro-kinetic cell allowed using sidewalls with the dimensions of a standard rectangular microscopy slide. This feature could be interesting for tweezing electrophoresis where it could facilitate a study of potentials of transparent Indium-Tin-Oxide (ITO) electrodes.

To demonstrate the flexibility resulting from all these features, we finish with a worked example for our electro-osmotic experiments. We here studied PS301 tracers in the presence of symmetric electro-osmotic flows realized with two walls of glass, PMMA or freshly prepared DMOAP coated glass. As expected, different electro-osmotic mobilities and coinciding electro-phoretic mobilities were observed. The previous electro-osmotic pumping experiments and micro-swimming experiments [25, 26, 50] were also conducted on these three substrates. The radial approach speed of the tracers towards the central IEX is shown in Fig. 11a for these three substrates. Far off the IEX the tracers are only dragged along by the electro-osmotic flow along the substrate. Closer to the IEX, the local field strength driving tracer electro-phoresis exceeds the average field strength driving electro-osmosis. Due to their own electro-phoretic motion the tracers slow and are finally halted and accumulated very close to the IEX [25, 26, 69]. Accordingly, the data in Fig. 11a are shown with two different symbols. Closed symbols for large and open symbols for short radial distances, respectively. In all cases, the radial dependence at large distances shows a power law decrease with an exponent close to -1. This is equivalent to a constant electro-osmotic solvent flux, *I*, through the wall of a cylinder of radius $r > r_{max}$ where r denotes the position of the velocity maximum. For each substrate this flux was calculated by averaging over all filled data points out to $r = 400\mu m$ and assuming a flow height of $10\mu m$.

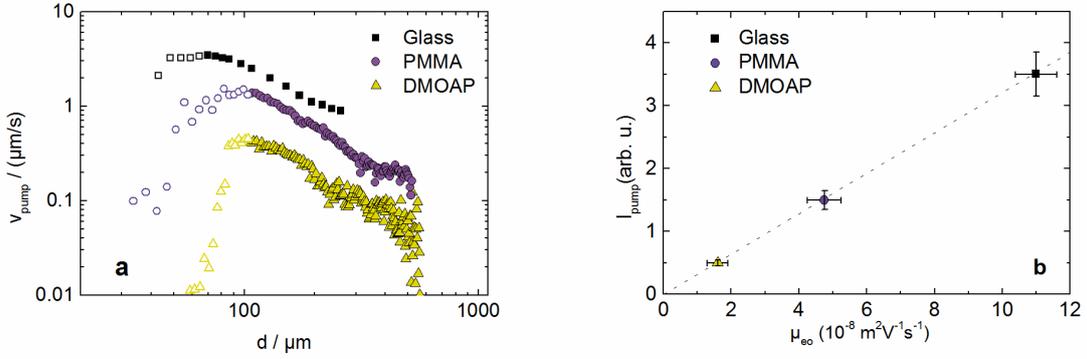

Fig. 11: Worked example: applying measured electro-osmotic mobilities to corresponding micro-fluidic pumping experiments. a) Radial dependence of tracer speeds measured in electro-osmotic pumping at a single ion exchanger bead of 45µm diameter for three different substrates: glass (squares), PMMA (circles) and DMOAP-coated glass close to the isoelectric point (triangles). Open symbols correspond to regions, where tracer motion is influenced by the electro-osmotic solvent flow *and* tracer elecectro-phoretic motion. Closed symbols denote data taken at larger distances, where tracers are moved *only* by the solvent flow. b) Electro-osmotic pump flux calculated from the data in a) in dependence on the electro-osmotic velocities determined in the present ISASH-LDV experiments using the cell with exchangeable side walls. The strictly linear dependence verifies the theoretical predictions of [25] and the analytical model proposed for modular swimming in [50].

A precise knowledge of the electro-osmotic mobilities now allows demonstrating that over the complete region denoted by filled symbols in Fig. 11a, the pump flux depends linearly on this quantity. This is shown in Fig. 11b and quantitatively confirms previous qualitative observations as well as theoretical expectation [25, 26]. It further shows that in all three experiments, the same distance averaged diffusio-electric field, $E_{global}$ drives the electro-osmotic flux. At any distance corresponding to a filled point $r \geq r_{max}$, $v_p = \mu_{eo} E_{global} / r$. Inside this radius the local diffusio-electric field becomes larger than its average value and the tracer velocity in addition becomes influenced by tracer electro-phoresis: $v_p = \mu_{eo} E_{global}./ r + \mu_{ep} E(r)$. Farniya et al. [68] have recently demonstrated, that from a variation of tracer mobility at assumed constant substrate mobility the spatial distribution of the diffusio-electric field for a catalytic pump could be derived. The present investigation is complementary as it provides data on a constant tracer mobility and differing substrate mobilities. Systematic experiments along this line to derive the field distribution for our electro-osmotic pump are under way and will be published elsewhere. However, even without exact experimental knowledge of involved fields, the present mobility data in combination with a precise experimental determination of the electrolyte concentration gradients [49] allow in depth modelling of the electro-osmotically driven solvent flows [25], a quantitative understanding of particle assembly by electro-osmotic pumps [26, 69] as well as testing theoretical concepts concerning modular micro-swimming [50, 70].

In conclusion, we used a recently introduced ISASH-LDV instrument in the integral configuration in combination with a novel electro-kinetic cell featuring exchangeable side walls. Precise and accurate values for electro-phoretic and electro-osmotic mobilities for different materials were obtained simultaneously and the possibility of using an internal mobility standard demonstrated. Performance and scope of this approach were discussed in detail. We anticipate that our versatile integral table-top approach will turn out to be very useful for flow measurements and electro-kinetic characterization in general and micro-fluidic application in particular.

## Conflicts of Interest

There are no conflicts of interest to declare.

## Acknowledgments

We are indebted to Ludmila Marotta Mapa, Holger Schweinfurth, Christopher Wittenberg and So Okomura for their contributions in developing ISASH-LDV. We like to thank all our theoretical colleagues, namely Raphael Roa, Marco Heinen, Angel Delgado, Christian Holm, Vladimir Lobashkin, Felix Carrique and Emilio Ruiz-Reina for the constant and encouraging interest in low-salt electro-kinetics and their support in mobility evaluation. Financial support of DFG (grant Pa-

459/18-1,2 within priority program SPP1726 "Microswimmers") and the Inneruniversitäre Forschungsförderung, JGU Mainz is gratefully acknowledged.

# References

1 A. Delgado, F. González-Caballero, R. J. Hunter, L. K. Koopal, J. Lyklema, *J. Colloid Interface Sci.*, 2007, **309**, 194

2 M. von Smoluchowski, *Bull. Acad. Sci. Cracovie, Classe Sci. Math. Natur*, 1903, **1**, 182

3 R.J. Hunter, *Zeta Potential in Colloid Science*, Academic Press, New York, 1981

4 Y. Yeh and H. Cummins, *Appl. Phys. Lett.*, 1964, **4**, 176

5 M. Deggelmann, Chr. Graf, M. Hagenbüchle, U. Hoss, Chr. Johner, HG. Kramer, Chr. Martin, R. Weber, J. Phys. Chem., 1994, **98**, 364

6 J. M. Roberts, J. J. O'Dea, and J. G. Osteryoung, *Anal. Chem.*, 1998, **70**, 3667-3673

7 J. F. Miller, *J. Colloid Interface Sci.*, 1992, **153**, 266

8 K. Schätzel, W. Wiese, A. Sobotta, M. Drewel, *J. Colloid Interface Sci.*, 1991, **143** (1), 287

9 L. G. Longsworth, D. A. McInnes, *Chem. Rev.*, 1939, **24**, 271

10 E. Uzgiris, *Rev. Sci. Instrum.*, 1974, **45**, 74

11 E. E. Uzgiris, *Prog. Surf. Sci.*, 1981, **10**, 53

12 K. Schätzel and J. Merz, *JCP,* 1984, **81**, 2482

13 A. Doren, J. Lemaitre, P. G. Rouxhet, *J. Colloid Interface Sci.*, 1989, **130**, 146

14 N. L. Burns, *J. Colloid Interface Sci.*, 1996, **183**, 249

15 N. L. Burns, K. Emoto, K. Holmberg, J. M. Van Alstine, and J. M. Harris, *Biomaterials*, 1998, **19**, 423

16 D. Botin, L. Mapa, H. Schweinfurth, B. Sieber, C. Wittenberg and T.Palberg, *J. Chem. Phys.*, 2017, **146**, 204904

17 A. Cohen, *Phys. Rev. Lett.*, 2005, **94** 118102

18 J. Palacci, B. Abécassis, C. Cottin-Bizonne, C. Ybert, and L. Bocquet, *Phys. Rev. Lett.*, 2010, **104**, 138302

19 J. Palacci, B. Abécassis, C. Cottin-Bizonne, C. Ybert, and L. Bocquet, *Soft Matter*, 2010, **8**, 980

20 I. Semenov, O. Otto, G. Stober, P. Papadopoulos, U. F. Keyser, and F. Kremer, *J. Colloid Interface Sci.*, 2009, **337**, 260

21 I. Semenov, S. Raafatnia, M. Sega, V. Lobaskin, C. Holm, and F. Kremer, *Phys. Rev. E*, 2013, **87**, 022302

22 G. S. Roberts, T. A. Wood, W. J. Frith, P. Bartlett, *J. Chem. Phys.*, 2007, **126**, 194503

23 F. Strubbe, F. Beunis, and K. Neyts, *J. Colloid Interface Sci.*, 2006, **301**, 302

24 Y. Hong, M. Diaz, U. M. Córdova-Figueroa, and A. Sen, *Adv. Funct. Mater.*, 2010, **20**, 1

25 R. Niu, P. Kreissl, A. T. Brown, G. Rempfer, D. Botin, C. Holm, T. Palberg, J. de Graaf, *Soft Matter*, 2017, **13**, 1505

26 R. Niu, E. C. Oğuz, H. Müller, A. Reinmüller, D. Botin, H. Löwen and T. Palberg, *PCCP*, 2017, **19**, 3104

27 J. L. Anderson, *Ann. Rev. Fluid Mech.*, 1989, **21**, 61

28 T. Palberg, H. Schweinfurth, T. Köller, H. Müller, H. J. Schöpe, A. Reinmüller, *Eur. Phys. J. Special Topics*, 2013, **222**, 2835

29 J. F. Miller, K. Schätzel, B. Vincent, *J. Colloid Interface Sci.*, 1991, **143**, 532

30 T. Palberg, H. Versmold, *J. Phys. Chem.*, 1989, **93**, 5296

31 T. Palberg, T. Köller, B. Sieber, H. Schweinfurth, H. Reiber, and G. Nägele. *J. Phys.: Condens. Matter*, 2012, **24**, 464109

32 M. Medebach, T. Palberg, *J. Phys. Condens. Matter*, 2004, **16**, 5653

33 H. Reiber, T. Köller, T. Palberg, F. Carrique, E. Ruiz-Reine and R. Piazza, *J. Colloid Interface Sci.*, 2007, **309**, 315

34 M. Medebach, L. Shapran, T. Palberg, *Colloid Surfaces B*, 2007, **56**, 210

35 J. Miller, K. Schätzel and B. Vincent, *J. Colloid interface Sci.*, 1991, **143**, 532

36 B. Berne, R. Pecora, *Dynamic Light Scattering*, John Wiley & Sons, Inc., New York, 1976

37 S. Komagata, *Researches Electrotech. Lab. Tokyo Comm.*, 1933, **348**, 1

38 T. E. Oliphant, Python for Scientific Computing, Comput Sci Eng, 2007, **9**, 10

39 A. Stipp, R. Biehl, Th. Preis, J. Liu, A. Barreira Fontecha, H. J. Schöpe, T. Palberg, *J. Phys. Condens. Matter*, 2004, **16**, S3885

40 D. Hessinger, M. Evers and T. Palberg, *Phys. Rev. E,* 2000, **61**, 5493

41 M. Medebach, R. Chuliá Jordán, H. Reiber, H.-J. Schöpe, R. Biehl, M. Evers, D. Hessinger, J. Olah, T. Palberg, E. Schönberger, P. Wette, *J. Chem. Phys.*, 2005, **123**, 104903

42 P. Wette, H.-J. Schöpe, R. Biehl, T. Palberg, *J. Chem. Phys.*, 2001, **114**, 7556

43 B. J. Kirby and E. F. Jr. Hasselbrink, *Electrophoresis,* 2004, **25**, 203

44 H. Falahati, L. Wong, L. Davarpanah, A. Garg, P. Schmitz and D. P. J. Barz, *Electrophoresis*, 2014, 35, 870

45 Y. Levin, *Rep. Prog. Phys.2002,* **65**, 1577.

46 G. Trefalt, T. Palberg, M. Borkovec, *Curr. Opn. Colloid Interface Sci.* 2017, **27**, 9

47 R. J. Messinger and T. M. Squires, *Phys. Rev. Lett.*, 2010, **105**, 144503

48 T. M. Squires, M. Z. Bazant, *J. Fluid Mech.*, 2004, **509**, 217

49 R. Niu, S. Khodorov, J. Weber, A. Reinmüller, T. Palberg, *New J Phys.*, 2017, **19**, 115014

50 R. Niu, D. Botin, A. Reinmüller, T. Palberg, *Langmuir*, 2017, **33**, 3450

51 T. Araki, H. Tanaka, *EPL*, 2008, **82**, 18004

52 D. Semwogerere, J. F. Morris, E. R. Weeks, *J. Fluid Mech.* 2007, **581**, 437.

53 D. A. J. Gillespie, J. E. Hallett, O. Elujoba, A. F. Che Hamzah, R. M. Richardson, and P. Bartlett, *Soft Matter*, 2014, **10**, 566

54 M. Castelain, F. Pignon, J.-M. Piau and A. Magnin, *J. Chem. Phys*, 2008, **128**, 135101

55 J. Lee, Z.-L. Zhou, G. Alas and S. H. Behrens, *Langmuir*, 2015, **31**, 11989

56 S. Stotz *J. Colloid Interface Sci.*, 1978, **65**, 118

57 M.Z. Bazant, M. S Kilic, B.D. Storey, A. Ajdari, *Adv. Colloid Interface Sci*, *2009*, **152**, 48

58 F. Strubbe, F. Beunis, T. Brans, M. Karvar, W. Woestenborghs, and K. Neyts, *Phys. Rev. X*, 2013, **3**, 021001;

59 J. E. Hallett, D. A. J. Gillespie, R. M. Richardson, and P. Bartlett, *Soft Matter*, 2018, **14**, 33

60 M. Evers, PhD thesis, Johannes Gutenberg Universität Mainz, 2000

61 M. Evers, N. Garbow, D. Hessinger, T. Palberg, *Phys. Rev. E*, 1998, **57**, 6774

62 V. Lobashkin, B. Dünweg, C. Holm, M. Medebach, T. Palberg, *Phys. Rev. Lett., 2007,* **98**, 176105

63 Á. V. Delgado, F. Carrique, R. Roa, E. Ruiz-Reina, *Curr. Opn. Colloid Interface Sci.*, 2016, **24,** 32

64 G. Trefalt, S.H. Behrens and M. Borkovec *Langmuir*, 2016, **32 (2),** 380

65 Y. Yeh and H. Z. Cummins, *Appl. Phys. Lett.*, 1964, **4**, 176

66 T. Palberg, M. Würth, *J. Phys I*, 1996, **6**, 237

67 S. Stotz, *J. Colloid Interface Sci.* 1978, **65**, 118

68 A. A. Farniya, M. J. Esplandiu, D. Reguera, and A. Bachtold, *Phys. Rev. Lett.*, 2013, **111**, 168301

69 R. Niu, T. Palberg, *Soft Matter* 2018, **14**, 3435

70 R. Niu, T. Palberg, *Soft Matter* 2018 Advanced Article DOI: 10.1039/c8sm00995c